\documentclass{elsart}
\usepackage{psfig}
\begin{document}

\begin{frontmatter}
\title{Numerical Observation of Disorder-Induced Anomalous Kinetics in the
$\mbox{A}+\mbox{A} \to \emptyset$ Reaction}

\author{Won Jae Chung and Michael W. Deem}
\address {Chemical Engineering Department, University of
California, Los Angeles, CA  90095-1592
}

\begin{abstract}
We address via numerical simulation the two-dimensional bimolecular 
annihilation reaction $\mbox{A} + \mbox{A} \to \emptyset$ in the 
presence of quenched, random impurities.  
Renormalization group calculations have
suggested that this reaction displays
anomalous kinetics at long times,
$c_{\rm A}(t) \sim at^{\delta -1}$,
 for certain types of
topological or charged species and impurities.
Both the
exponent and the prefactor 
depend on the strength of disorder.  The decay
exponents determined from our simulations agree well with the values
predicted by theory.  The observed renormalization of the prefactor also
agrees well with the values predicted by theory.  
\end{abstract}
\end{frontmatter}



\section{Introduction}

Surface reactions in two-dimensions show a variety of
interesting behavior.  Qualitatively, the  difference between
two and three dimensions is that diffusive mixing is less
effective in two (and one) dimensions.  With less mixing, transport
limitations become more significant, and this leads to a breakdown of
the law of mass action in two dimensions.
Formally, two-dimensions is the upper critical 
dimension for bimolecular surface reactions \cite{Peliti,Lee2}.
  The law of mass action, or local chemical kinetics, is
the simplest theory with which to predict the reactant concentration in the 
$\mbox{A} + \mbox{A}
~{\mathrel{\mathop{\to}\limits^{k}_{}}}~
\emptyset$
 reaction:
\begin{eqnarray} 
\frac{d c_{\rm A}}{d t} &=& -k c_{\rm A}^2 
\nonumber \\
c_{\rm A}(0) &=& n_0 \ , 
\label{1} 
\end{eqnarray}
where $k$ is the reaction rate constant.
This theory
fails because, without input reactants, the reaction is diffusion limited
at long times.  The approach of Eq.\ (\ref{1}) does not take into account
diffusion limitations, and so this theory cannot be a predictive one.
If, in addition, the reaction occurs in a disordered medium, there will
be an interplay between the effects of disorder and
diffusion limitations, and the simple law of
mass action can make no statements about the resulting
kinetics.
Finally, the law of mass action
cannot account for any clustering or
other collective behavior 
of the reactants, since it effectively assumes that
the reactants are
perfectly well mixed.

A more sophisticated level of theory for chemical reactions is 
the reaction
diffusion partial differential equation:
\begin{eqnarray} 
\frac{\partial c_{\rm A}}{\partial t} &=& D \nabla^2 c_{\rm A} + \beta D \nabla 
\cdot \left(c_{\rm A }\nabla v\right) - k c_{\rm A}^2  \ .
\label{2} \end{eqnarray}
We will be interested in the case of random initial conditions, where
$c_{\rm A}({\bf x}, 0)$ is a Poisson  random number with average
$n_0$.  
Here $D$ is the diffusivity, $v({\bf r})$ is the quenched potential in which
the reactants diffuse, and $\beta = 1/(k_{\rm B} T)$ is
the inverse temperature.  This approach, surprisingly, is 
also qualitatively wrong in two dimensions.  This approach fails because it
is a type of mean field theory, and two dimensions is
the upper critical dimension for
bimolecular kinetics, below which mean field theory fails.
If one insists to use Eq.\ (\ref{2}), one could predict
the true behavior only
by using a renormalized, effective reaction rate in place of the
bare rate.  This effective reaction rate is exactly what  the
renormalization group treatment of this problem predicts 
\cite{Peliti,Lee1,Deem1,Deem2}.  What does the reaction diffusion equation
leave out, which causes it to miss the renormalization of the effective
reaction rate?
  Technically this approach
does not capture the integral occupation number constraint at each site.
 That is, at each place in space, there can be only an 
integer number of reactants.  This seemingly trivial constraint
is what leads to a renormalization of the effective
reaction rate.

Previous simulation work has been conducted for a variety of
simple reactions in two dimensions.  Many simulations have been performed on
clean metal surfaces (for a review
see \cite{Weinberg}).
 Simple systems, such as
oxidation of CO on single crystal Pt(110), show
surprisingly rich behavior \cite{Graham}, ranging
from spirals and standing waves to chemical
turbulence \cite{Ertl,Jakubith}.
Simulation studies have confirmed several analytical
results for the simpler systems (see, for example, \cite{Kopelman2}).
Fascinating effects have been observed on surfaces with patterned
disorder \cite{Kevrekidis1,Kevrekidis2}.
Surfaces with spatially random adsorption energies have been
shown to lead to a variety of
phases observed at steady state \cite{Redner2}.
Finally, simulations 
  have been extended to the case of
fractal media (see, for example, \cite{Zumofen,Kopelman}).

Unlike these previous simulation studies, we are interested in the kinetics
of the ionic reaction $\mbox{A}^+ + \mbox{B}^- \to \emptyset$ in
two dimensions in the presence of quenched, charged disorder.
The disorder, and the reactants themselves, can either be simple
electrostatic charges or topological ``charges'' such as dislocations
or disclinations.  For either system, the quenched defects interact
with the diffusing reactants via a long-ranged, Coulombic potential
energy.  The upper critical dimension for this
diffusive process is also two.  The effects of this disorder
are dramatic, and, in fact, the potential energy
landscape induced by the random charges is so rugged that
the motion is sub-diffusive in two dimensions
\cite{Fisher,Kravtsov1}.  
In other words, this special type of
disorder causes the mean square displacement of random walkers
to increase sub-linearly
with time, $\langle r^2(t) \rangle \sim b t^{1-\delta}$.  The
scaling exponent is continuously variable in the strength
of disorder and can be found exactly 
\cite{Kravtsov2,Bouchaud1,Bouchaud2,Honkonen1,Honkonen2,Derkachov1,Derkachov2}.

Each of the reactions
$\mbox{A} + \mbox{A} \to \emptyset$,
$\mbox{A} + \mbox{B} \to \emptyset$, and
$\mbox{A}^+ + \mbox{B}^- \rightleftharpoons \mbox{AB}$
have been studied
in the presence
of quenched, singular disorder by field-theoretic
methods \cite{Deem1,Deem2,Deem3}.
  Renormalization group theory was used to derive
the long-time values for the reactant concentrations.
Interestingly, the $\mbox{A}^+ + \mbox{B}^- \rightleftharpoons \mbox{AB}$ ionic reaction
at high temperature behaves like the $\mbox{A} + \mbox{A} \to \emptyset$
model when the back reaction rate is equal to zero  \cite{Deem3,Toussaint}.
This is because the Coulomb interaction between the reactants inhibits
the reactant segregation that would otherwise occur in this
reaction.
We chose to study
the $\mbox{A} + \mbox{A} \to \emptyset$ reaction as a
simplified model.

We can understand the results of these renormalization group studies
by a simple argument.
The $\mbox{A} + \mbox{A} \to \emptyset$ reaction becomes diffusion limited
at long times in the quenched disorder.  Simple theory for diffusion-limited 
reactions, therefore,
 predicts that the effective rate constant will be
proportional to the diffusivity.  Since the diffusion is anomalous in this
disorder, $\langle r^2 (t) \rangle \sim 4 D t (t/t_0)^{-\delta}$, we should
expect $k_{\rm eff}
\propto D_{\rm eff} \sim D (t/t_0)^{-\delta}$, where $\delta$ is the same
measure of the
strength of disorder as above.  Here $D$ is the bare diffusivity, and
$t_0 = (\Delta r)^2/(2 D)$ is a characteristic time for diffusion
on a lattice of spacing $\Delta r$.
  We then might expect to use mean field theory with
this effective reaction rate to predict the concentration of reactants
\begin{equation} 
c_{\rm A} (t) \sim \frac{1}{k_{\rm eff} t} \sim
\frac{1}{k^* t} \left(\frac{t}{t_0}\right)^\delta
~~~{\rm as}~
t \to \infty \ ,
\label{3} 
\end{equation}
where $k^*$ is some fixed-point effective reaction rate to be determined
by a more careful calculation.
This simple argument is in agreement with rigorous renormalization
group results, which predict \cite{Deem1}
\begin{equation}
k^* = 3 \beta^2 \gamma D  \ .
\label{3a}
\end{equation}

We here test these renormalization group predictions.
We analyze via numerical simulation the behavior of the 
$\mbox{A}+\mbox{A} \to \emptyset$ reaction in two-dimensions in the 
presence of singular disorder.  
We calculate by simulation the concentration profiles at long 
times for various strengths of disorder.
We compare both the exponent and the prefactor predicted by
renormalization group arguments against those measured via simulation
\cite{Deem1}.
This paper is organized as follows.  In Section 2
we describe the disorder that is appropriate
for ionic systems.  We
express the effects of the quenched disorder in terms
of a random potential.
  In Section 3, we describe the master equation that defines
our model of the
reaction.  We describe a method for numerically solving the master equation 
by a Poisson process that is
effective for strong disorder.  In Section 4, we take an alternative approach
and describe an exact stochastic partial differential 
equation (SPDE) that can be derived by field-theoretic means from the master
equation.    We describe a method for solving the SPDE numerically
that is effective for weak disorder.
In Section 5, we implement these solution techniques
and analyze the long-time behavior of the  reactant concentration
at various strengths of 
disorder.  We discussion our numerical results  in light of the theoretical
predictions in Section 6.  We conclude in Section 7.

\section{The Disorder}
The two-dimensional ionic reaction $\mbox{A}^+ +\mbox{B}^- 
\to \emptyset$ behaves like 
the nonionic bimolecular annihilation reaction $\mbox{A} +\mbox{A} \to 
\emptyset$ because the Coulomb interaction inhibits the reactant 
segregation that naturally occurs in the $\mbox{A} +
\mbox{B} \to \emptyset$ 
 reaction \cite{Deem3,Toussaint}.  
This, then, implies that the reactant concentration
for the
$\mbox{A}^+ + \mbox{B}^- \to \emptyset$ model is
proportional to that for the $\mbox{A} + \mbox{A} \to \emptyset$
model at long times.   The presence of additional, quenched, charged disorder
does not change this argument.
The $\mbox{A} +\mbox{A} \to \emptyset$ reaction is much easier to
simulate than the ionic $\mbox{A}^+ +\mbox{B}^- \to \emptyset$ reaction,
since one does not have to track the long-range forces between
the reactants.
For this reason, we choose to
simulate the $\mbox{A} +\mbox{A} \to \emptyset$ model
as a general test of the renormalization group predictions
for two-dimensional
bimolecular reactions in the presence of singular disorder
\cite{Deem1,Deem2,Deem3}.

The type of physical systems that we have in mind are systems with 
``ionic'' disorders.  Such disorder could occur, for example,
 on the surface of an ionic 
crystalline lattice.  Imagine that
the lattice has line defects caused by 
dislocation line pairs forming line vacancies or line 
interstitials.  If these defects were immobile, they would
generate a random, 
quenched electrostatic potential on the surface.  The
interaction between ionic reactants on the surface
and these line charges would
be logarithmic, which is technically
``relevant'' in two dimensions.
The 1/r interactions between the ions themselves is technically
``irrelevant'' and can be ignored.

Alternatively, our system can be viewed as modeling the dynamics of
a collection of line dislocations in  a
solid.  The hexatic-solid transition in two dimensional monolayers
occurs by pairing of dislocations  and is a good
example of an ionic reaction, albeit with vector ``charges'' \cite{NelsonII}.
If there were some additional dislocations pinned by defects,
these defects would interact  logarithmically with the moving dislocations,
just as in our model.  A physical system that more closely resembles
our model is the liquid-hexatic transition in two dimensional fluids.
Here the transition occurs by pairing of disclinations, which are
a perfect manifestation of topological charges that interact logarithmically
\cite{NelsonII}.  Additional, pinned disclinations would generate the
quenched disorder present in our model.

Experiments have, indeed, seen effects of pinned disorder on the
liquid-hexatic transition for both flux lines in charged density wave
systems \cite{LieberI,LieberII} and surfactants in hexatic multilayers
in Langmuir-Blodgett films
\cite{Zasadzinski}.  
Detailed renormalization
group calculations are in agreement with the observed effects
\cite{Deem3}.
 An additional physical system, for which
experiments could be performed, are charged vortices in a thin fluid
film between two spatially-addressable electrodes \cite{Tabeling}.

We use a quenched random potential 
to represent the potential induced by the immobile defects.
 The random potential is Gaussian at long
wavelengths \cite{Deem0}, and the appropriate 
form of the potential-potential correlation function at long wavelengths is
\begin{equation}
\hat \chi_{vv}(k) \sim \gamma/k^2 \ .
\label{4}
\end{equation}
Here $\gamma$ is a measure of the strength of disorder. It
is roughly proportional to the square of the density of defects.
Our simulation takes place on a square lattice, and
 a natural 
form of the potential-potential correlation function  is
\begin{equation}
\hat \chi_{vv}(k_x,k_y) = \frac{\gamma (\Delta r)^2}{4 - 2\cos(k_x \Delta r)
- 2 \cos(k_y \Delta r)} \ ,
\label{5}
\end{equation}
where $\Delta r$ is the lattice spacing.
This form reduces to that of Eq.\ (\ref{4}) at long wavelengths,
$k \to 0$, and in the
limit of small lattice spacing, $\Delta r \to 0$.
We generate the random potential $v({\bf r})$ by first generating
$\hat v({\bf k})$ in Fourier space, where the values at different
wave vectors are independent Gaussian random numbers,
 and then performing an inverse fast
Fourier transform \cite{Deem0a,Deem5}.
We use periodic boundary conditions for the random lattice.

A random potential of this form is known to
have a significant effect 
on diffusing species, resulting in an anomalous mean square
displacement at long times 
\cite{Fisher,Kravtsov1,Kravtsov2,Bouchaud1,Bouchaud2,Honkonen1,Honkonen2,Derkachov1,Derkachov2}
\begin{equation}
\left\langle r^2(t) \right\rangle \sim 4 D t \left(
\frac{t}{t_0}
\right)^{-\delta} \ .
\label{6}
\end{equation}
The scaling exponent, $1-\delta$, depends on the strength of disorder.  For
Gaussian disorder,
the exponent is exactly given by
\begin{equation}
\delta = \left[1 + \frac{8 \pi}{\beta^2 \gamma} \right]^{-1} \ .
\label{7}
\end{equation}

\section{The $\mbox{A}+\mbox{A} \to \emptyset$ Reaction}
Our model of the $\mbox{A}+\mbox{A} \to \emptyset$ reaction
can be defined mathematically
by a master equation.
We assume that the reaction
takes place on a square $N \times N$ 
lattice with lattice spacing
$\Delta r$. The 
reactants move diffusively in a random
potential on the lattice.  The reactants
annihilate at rate $k$ whenever two meet 
on the same lattice site.  
The master equation describes how  the probability of
each possible configuration of particles on the lattice,
$P(\{n_i\},t)$, 
changes with time.  
 The appropriate master equation for our system is
\begin{eqnarray}
\frac{\partial P(\{n_i\},t)}{\partial t} &=& \sum_{ij}
[\tau_{ji}^{-1} (n_j +1)P(...,n_i -1,n_j +1,...,t) 
- \tau_{ij}^{-1} n_i P] 
\nonumber \\
&&+ \frac{k}{2(\Delta r)^2} \sum_i [(n_i+2)(n_i +1)
 P (...,n_i +2,...,t) 
\nonumber \\
&&~~~~~~~~~~~~~~~~~~-n_i (n_i -1)P] \,
\label{8}
\end{eqnarray}
where $n_i$ is the number of particles at site $i$, $D$ is the diffusivity, 
$k$ is the reaction rate constant, and $\Delta r$ is the lattice spacing.
Here $\tau_{ij}^{-1}$, to be defined below,
 is the rate at which particles hop from lattice site $i$ to
nearest neighbor lattice site $j$.
The 
summation over $i$ is over all sites on the lattice and the
summation over $j$ is over all nearest neighbors of site $i$. We
place the reactants randomly on the lattice at time $t=0$.  The
initial concentration at any given site will, thus, be 
Poisson, with average density that we define to be $n_0$:
\begin{eqnarray}
P(n_i) &=& 
\frac{[n_0 (\Delta r)^2]^{n_i}}{n_i !}
{\rm e}^{-n_0 (\Delta r)^2} 
\nonumber \\
\langle n_i /(\Delta r)^2 \rangle &=& n_0 \ ,
\label{9}
\end{eqnarray}

This master equation can be solved exactly by directly considering the particle 
reaction and diffusion process \cite{Doering}.  
The master equation implies that a given configuration of reactants on the
lattice follows a
continuous-time, Markovian, Poisson process.  In this process,
all possible changes to the lattice configuration occur at
 specific rates.  That is, both the reaction and diffusion moves
have rates.
The rates of diffusion depend on the values of the  potential  on the
lattice.  Certain moves are favored over others, due to local
gradients in the
potential.  The rate of hopping to a nearest neighbor site
is $\tau_{ij}^{-1}$.  There are four of these rates at each lattice site.
  The rate for hopping to the right is
\begin{equation}
\tau_{\rm R}^{-1}(x,y) = \frac{D}{(\Delta r)^2} \exp\left\{\beta
\left[v(x,y)-v(x+\Delta r,y) \right]/2\right\} \ ,
\label{10}
\end{equation}
for hopping to the left is
\begin{equation}
\tau_{\rm L}^{-1}(x,y) = \frac{D}{(\Delta r)^2} \exp\left\{\beta
\left[v(x,y)-v(x-\Delta r,y) \right]/2\right\} \ ,
\label{11}
\end{equation}
for hopping above is
\begin{equation}
\tau_{\rm U}^{-1}(x,y) = \frac{D}{(\Delta r)^2} \exp\left\{\beta
\left[v(x,y)-v(x,y+\Delta r) \right]/2\right\} \ ,
\label{12}
\end{equation}
and for hopping below is
\begin{equation}
\tau_{\rm D}^{-1}(x,y) = \frac{D}{(\Delta r)^2} \exp\left\{\beta
\left[v(x,y)-v(x,y-\Delta r) \right]/2\right\} \ .
\label{13}
\end{equation}
At each lattice site, there are $n_i(n_i-1)/2$ possible reaction moves.
Each of these moves has the rate 
\begin{equation}
\tau_{\rm rxn}^{-1} = \frac{k}{(\Delta r)^2} \ .
\label{13a}
\end{equation}
Overall, for the whole lattice
there are $4 \sum_i n_i$ possible diffusion moves
and $\sum_i
n_i (n_i -1)/2$ possible reaction moves.

A Markov process is initiated by placing the particles randomly
on the surface, with average density $n_0$.
The Markov chain is then generated by picking one of these
possible events according to its probability of occurrence and 
incrementing time according to its rate of occurrence.
The probability of event $\alpha$ occurring, out of all the possible
diffusion and reaction moves, is
\begin{equation}
P({\rm event~ }\alpha) = \frac{\tau_\alpha ^{-1}}{\sum_\gamma \tau_\gamma
^{-1}} \ .
\label{15}
\end{equation}
Since the process is Poisson, time is incremented by 
\begin{equation}
\Delta t = \frac{-\log \zeta}{\sum_\gamma \tau_\gamma^{-1}} \ ,
\label{16}
\end{equation}
where $\zeta$ is a uniformly distributed random number between zero and 
one.  
This step of the Markov chain
is repeated until only zero or one particles remain on the lattice.

We find this procedure to be efficient when the strength of disorder,
$\beta^2 \gamma$, is large.
For computational convenience, we make two approximations.
First,
when a particle moves to a site that is already occupied by 
another particle, the annihilation reaction occurs immediately.  In other
words, we set the reaction rate, $k$, equal to infinity.
Note that the renormalization group predictions are independent of the
bare reaction rate \cite{Deem1}.
Physically this is because
the reaction is diffusion-limited at long times, and so
the bare reaction rate does not matter.
  Second,
we assume that
each particle has an equal probability of being moved
\begin{equation}
P({\rm moving~particle} ~ \alpha) = \frac{1}{n} \ ,
\label{15a}
\end{equation}
where n is the total number of particles on the lattice.  The
direction of movement of a chosen particle, however, is specified
by the local gradient of the random potential, as in
Eqs.\ (\ref{10}-\ref{13}).
Specifically, we
generate another uniformly distributed random number, which is
compared against the four 
different hop probabilities:
\begin{equation}
P({\rm hop} ~ i) = \frac{\tau_i^{-1}}{\sum_j \tau_j^{-1}} \ .
\label{16a}
\end{equation}
The chosen move is performed, and time is incremented 
by 
$\Delta t = N^2/(n \sum_{ij} \tau_{ij}^{-1})$.
The uniform choice of reactant to move and the approximate time incrementation
are both exact in the long-time limit when $\Delta r \to 0$ in the
hopping rates (\ref{10})-(\ref{13}).

\section{The Stochastic Equation}
Alternately, we can derive a stochastic partial differential equation by mapping the 
master equation onto a field theory.  The field theory looks like 
a Bosonic quantum field theory due to the constraint of
an integral occupation number at each lattice site.  
The continuous-time stochastic partial differential
equation for the $\mbox{A}+\mbox{A} 
\to \emptyset$ reaction is \cite{Deem1}
\begin{eqnarray}
\frac{\partial a}{\partial t} &=& D \nabla^2 a + \beta D \nabla \cdot 
\left( a \nabla v\right) - k a^2
 +\mathrm{i} \eta a
\nonumber \\
a({\bf x}, 0) &=& n_0 \ ,
\label{17}
\end{eqnarray}
where the real, Gaussian random field $\eta({\bf x},t)$ has zero mean and 
variance $\langle \eta({\bf x},t) \eta({\bf x}',t') \rangle = k 
\delta({\bf x}- {\bf x}') \delta(t-t')$.  The physical concentration 
is given by taking the average of the solution over the random field 
$\eta$.  Since the reaction occurs on a lattice,
the spatial derivative symbols are actually
a shorthand for
finite difference  expressions.  The field-theoretic derivation
shows that this stochastic differential equation should be interpreted
in the It{\"o} sense.

The stochastic partial differential equation was solved
numerically by using a second-order, semi-implicit method.
 This 
method gives exact solutions to the master equation
without any approximations, in the limit that the integration
time step tends to zero.
The process of solving the stochastic partial differential equation
takes place on a square lattice.
  We define $b[a(x,y,t)] = D \nabla^2 a - k a^2$ and
$\mu = \sqrt k / \Delta r$.  An
explicit method for Eq.\ (\ref{17}) 
equation is given by \cite{Petersen}
\begin{eqnarray}
a(x,y,t+\Delta t) &=& a(x,y,t) 
\nonumber \\
&&+ \frac{\Delta t}{2}
\bigg\{ b[a(x,y,t)] + b\bigg[a(x,y,t)
+ \mathrm{i} \mu \sqrt t a(x,y,t)z(x,y,t) \nonumber \\
&&~~~~~~~~ + \Delta t b[a(x,y,t)]
\bigg]\bigg\}
\nonumber \\
&&+ \mathrm{i} \mu \sqrt t \left[a(x,y,t)
 + \frac{\Delta t}{2} b[a(x,y,t)] \right]
z(x,y,t)
\nonumber \\
&&- \frac{\Delta t}{2} \mu^2 a(x,y,t) [ z(x,y,t)^2-1] \ .
\label{14}
\end{eqnarray}
Here $z(x,y,t)$ is a Gaussian random number of unit variance.
These random numbers are uncorrelated at different positions and times.
This numerical method is accurate to $O[(\Delta t)^2]$ in the weak sense.
A semi-implicit method is generated by replacing 
$b\left[a + \mathrm{i} \mu \sqrt t  a z + \Delta t b(a) \right]$ on
the right hand side with
$b[a(x,y,t+\Delta t)]$.
This semi-implicit method is also accurate to $O[(\Delta t)^2]$ in the
 weak sense.
Direct implementation of
this approach leads to a complicated, non-linear
matrix equation to solve.  We use, instead, an operator-splitting approach
\cite{Press}.  On even integration time steps, we use
a method implicit in the $x$-diffusion terms
and explicit in the $y$-diffusion and
reaction terms.  On odd integration time steps, we use
a method implicit in the $y$-diffusion terms
and explicit in the $x$-diffusion and
reaction terms.   This operator-splitting, semi-implicit approach
leads to tridiagonal matrix equations that are simple to solve.
 We found that including a time step implicit in the reaction
terms was not useful.
The periodic boundary conditions enter these equations
through the finite-difference definition of the
diffusion terms.

\section{Results}
For weak disorder,
$\beta^2 \gamma < 1$, the master equation was solved by
numerically integrating the exact stochastic differential equation.  
The concentrations produced in a typical simulation run
are illustrated in Figure 1.
\begin{figure}[t]
\centering
\leavevmode
\psfig{file=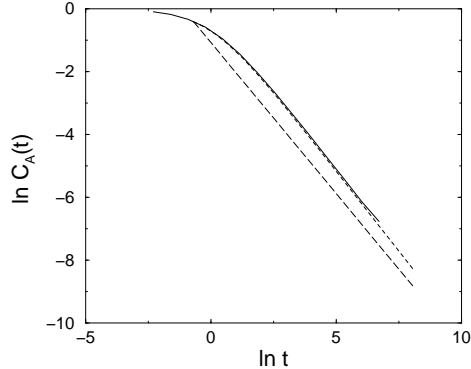,height=2in,angle=-90}
\caption[]
{\label{fig1}
The reactant concentration as a function of time for $\beta ^2
\gamma = 1$, $k=1$, and $n_0=1$.  The simulation data are shown by the
solid line.  The long-time value predicted by renormalization group
arguments is shown by the long-dashed line.  The concentration predicted by
renormalization group calculations for all times is shown by the 
short-dashed line.  At times longer than those shown here, the two dashed
lines converge.
}
\end{figure}
The simulation values at short times are quite far off from
the long-time values predicted by the renormalization group treatment.
This is because the asymptotic scaling of the
reaction concentration is reached only for very long times
for the parameters of Figure 1.
At very long times,
for a suficiently large lattice,
 the concentration observed by simulation
would be the asymptotic value predicted
by theory.
 To achieve an agreement between theory and simulation, we also show in
Figure 1 the concentration predicted for all times with the
use of the running coupling in Eq.\ (\ref{1}).  This approximate
concentration profile is derived in a straight-forward fashion from
the flow equations for this system \cite{Deem1}:
\begin{eqnarray}
c(t) &=& 
\frac{
(t/t_0)^{\delta -1}
}{n_0
(t/t_0)
^{\delta -1} + k(l^*)t_0} 
\nonumber \\
k(l^*) &=& \frac{3 \beta ^2 \gamma D}{1+ (3 \beta ^2
\gamma D/k_0 -1 )
(t/t_0)^{-3 \delta}} \ .
\label{20}
\end{eqnarray}
As we can see from this more detailed theoretical prediction, 
the renormalization of the effective reaction rate to the
asymptotic value is rather slow, occurring as $(t/t_0)^{-3 \delta}$.
At long times,
the level of noise in the simulation data becomes more significant.
This is because there are fewer particles at longer times, and
so the statistical error is greater.
If a greater lattice size or a smaller time step were used, then less
noise would be observed at these times.  Balancing these considerations
against computational feasibility,  we chose to use
$2048 \times 2048$ square lattices.

The slow renormalization of the effective reaction rate prevents
a direct simulation of the reaction for arbitrary values of the
parameters.
To counteract the effect of the slow renormalization,
we performed additional
simulations by starting with the renormalized reaction
rate predicted from theory, $k = k^* = 3 \beta^2 \gamma D$.
For runs with this initial value of the reaction rate, the concentrations
reach the asymptotic scaling at fairly short times.

To determine the observed values of the
the decay exponent, $1- \delta^*$, 
and
the renormalized reaction rate, $k^*$,
a log-log plot is made for the
concentration of species A versus time.  The slope of this plot
gives us the renormalized decay exponent, and the y-intercept
gives us the renormalized reaction rate.  With this method,
far more data points will be plotted and used at long times than
intermediate times in the
fit to determine the renormalized
values.  Therefore,
the values determined from the fit
will be heavily biased by data points gathered
at long times.  This biasing is artificial, since the concentration
profile must be continuous as a function of time, and so we should not
overweight the long-time section.
  To eliminate this bias, we used
data points in powers of two to perform the fits.  For example,
data points corresponding to times  steps
1, 2, 4, 8, 16, 32, 64, \ldots\  were used to determine the slope and the
prefactor.    Note that to determine $k^*$ we need to assume a value
for $t_0$.  We use $t_0 = (\Delta r)^2 / (2 D)$, which is suggested by
simple perturbation theory for this problem.

Figure 2 compares the exponent of the concentration decay observed by
simulation with that predicted by theory.
\begin{figure}[t]
\centering
\leavevmode
\psfig{file=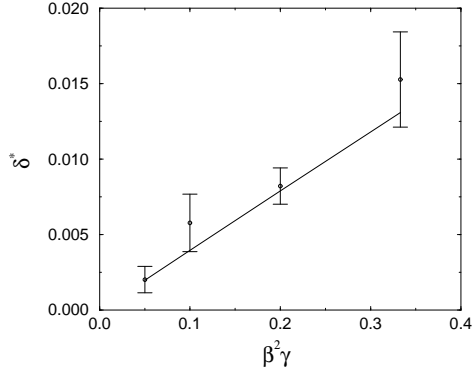,height=2in,angle=-90}
\caption[]
{\label{fig2}
Presented are the observed values for the exponent of the
reaction concentration decay, determined from 
$c_{\rm A}(t) \sim a t^{\delta^*-1}$,
as a function of the strength of disorder.
Renormalization group predictions are shown by the solid line.
}
\end{figure}
Figure 3 compares the effective reaction rate
observed by simulation with that predicted by theory.
\begin{figure}[t]
\centering
\leavevmode
\psfig{file=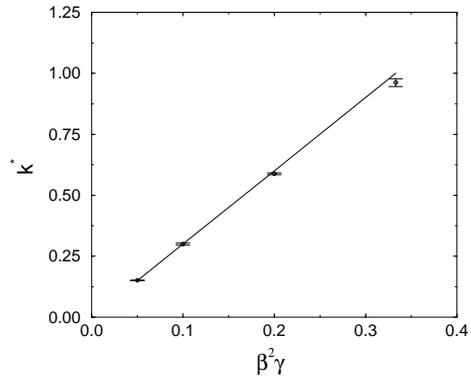,height=2in,angle=-90}
\caption[]
{\label{fig3}
Presented are the observed values of the renormalized
reaction rate, determined from 
$c_{\rm A}(t) \sim (t/t_0)^{\delta^*} / (k^* t)$,
as a function of the strength of disorder.
Renormalization group predictions are shown by the solid line.
}
\end{figure}
 These two plots show that the
theoretical values agree well with those obtained by simulation.
The error bars were determined by considering both random and systematic errors.  We
found that the most conservative method 
was to vary the smallest time at which we start the fit.
We found that the slope and prefactor were
somewhat sensitive to which data points were used in the calculations.  
The error bars encompass both the random, statistical errors and the
systematic error associated with the range of data that we fit.

For strong disorder,
$\beta^2 \gamma > 1$, the approximate Poisson process was used to solve 
the master equation.  For each set of parameters, we performed 10 simulations
on $2048 \times 2048$ lattices to generate average concentration
profiles. As before, a $\log c_{\rm A}$ versus $\log t$ plot  of the
simulation data was used 
to determine  the
decay exponent, $1- \delta^*$,
and the renormalized reaction rate, $k^*$.
   Also as before, we used data
exponentially spaced in time.  For strong disorder,
the renormalization of the effective reaction rate constant occurs
very quickly.  

Figure 4 compares the exponent of the concentration decay observed by
simulation with that predicted by theory.
\begin{figure}[t]
\centering
\leavevmode
\psfig{file=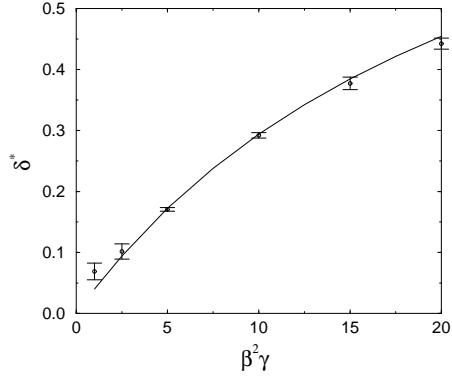,height=2in,angle=-90}
\caption[]
{\label{fig4}
Presented are the observed values for the exponent of the
reaction concentration decay, determined from 
$c_{\rm A}(t) \sim a t^{\delta^*-1}$,
as a function of the strength of disorder.
Renormalization group predictions are shown by the solid line.
}
\end{figure}
Figure 5 compares the effective reaction rate
observed by simulation with that predicted by theory.
\begin{figure}[t]
\centering
\leavevmode
\psfig{file=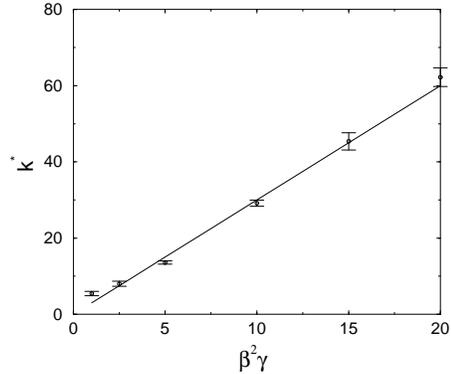,height=2in,angle=-90}
\caption[]
{\label{fig5}
Presented are the observed values of the renormalized
reaction rate, determined from 
$c_{\rm A}(t) \sim (t/t_0)^{\delta^*} / (k^* t)$,
as a function of the strength of disorder.
Renormalization group predictions are shown by the solid line.
}
\end{figure}

\section{Discussion}

The most significant prediction of the renormalization group studies
is the decay exponent, $c_{\rm A}(t) \sim a t^{1-\delta}$, where 
$\delta$ is given by Eq.\ (\ref{7}).  All simulations for weak disorder,
$\beta^2 \gamma < 1$, show an observed slope that is  in agreement
with this prediction, to within the error bars.   
This is significant,
since the renormalization group studies are, in principle, expansions
in the parameter $\beta^2 \gamma$, and so the predictions are 
strictly accurate only for weak disorder.  
It was anticipated 
in the renormalization group
treatment, however, 
that the result for the
decay exponent may be exact to all orders of $\beta^2 \gamma$ \cite{Deem1}.

For strong disorder, $\beta^2 \gamma > 1$, some small discrepancies
exist between the simulation and the theoretical predictions for the
decay exponent.  The approximate Poisson process method  that
we used to solve the
master equation requires the effective reaction rate to renormalize
from infinity to $k^*$.  This process occurs slowly for weak
disorder, as shown by
 Eq.\ (\ref{20}), and this is the reason for the discrepancy for 
$\beta^2 \gamma \approx 1$.  Simulations on lattices larger than
$2048 \times 2048$, which would
provide data to lower concentrations, would eventually yield an
exponent in agreement with the predicted value.

The renormalization group studies also predict the fixed-point value
of the effective reaction rate, Eq.\ (\ref{3a}).  This prediction,
strictly speaking, is an expansion in the disorder strength, and we
would generally expect there to be corrections higher order in
$\beta^2 \gamma$.  All of the simulations
for weak disorder are in agreement with this prediction.
For $\beta^2 \gamma \approx 0.33$, there is a small discrepancy not
contained within the error bars.
The numerical integration of Eq. (\ref{17}) is exceedingly difficult
for large $\beta^2 \gamma$, and the  small discrepancy
between simulation and theory
is most likely due to use of a time step that was not quite small enough.
Resolution of this discrepancy would have required 
an integration time step substantially smaller than is feasible, even on our
high-performance workstations.

Even for strong disorder, the simulations show an effective reaction
rate in agreement with the predicted values.  This is significant on 
two counts.  First, the renormalization group predictions are only
a first order approximation for $k^*$, and it is surprising that the
higher order corrections are so small for $\beta^2 \gamma \approx 20$.
Second, the simulation technique for strong disorder requires the
effective reaction  rate to renormalize from infinity to a finite
$k^*$.  That this happens so effectively is impressive.  
 The discrepancies in the observed
effective reaction rate for $\beta^2 \gamma \approx 1$ are, again, due to the
slow renormalization of the effective reaction rate.

\section{Conclusions}
Our numerical simulation of the 
$\mbox{A} + \mbox{A}
~{\mathrel{\mathop{\to}\limits^{k}_{}}}~
\emptyset$ reaction
is in agreement with the field-theoretic renormalization
group predictions \cite{Deem1}. 
An anomalous
decay exponent, predicted by
the renormalization group studies, is  observed.  No significant
discrepancy between the numerical and analytical predictions are
observed over a wide range of disorder strengths.  More
impressively, the effective reaction rate observed from the numerical
simulations is in quantitative agreement with the renormalization
group predictions for the same wide range of disorder strengths.
This agreement is unexpected and may signify that the one-loop
renormalization group results are more accurate than anticipated.
With these simulations, 
we now have satisfying agreement between rigorous field-theoretic
results, simple physical  arguments, and exact numerical results
for this interesting case of anomalous kinetics.

\section*{Acknowledgments}
This research was supported by the National Science Foundation
through grants CHE--9705165 and CTS--9702403.

\bibliography{won}

\end{document}